%% file: main.tex
\newif\ifjournal
\newif\ifreport
\newif\ifexperiments

\ifjournal
  \documentclass[lettersize,report]{IEEEtran}
\else\ifreport
  \documentclass{ornltm}
  \usepackage[authordate,strict,backend=biber,autolang=other,bibencoding=inputenc]{biblatex-chicago} 
  \bibliography{main}
\else
  \documentclass[twoside,leqno,twocolumn]{article}

  \usepackage{enumitem}
  \setlist{noitemsep}

  \usepackage{orcidlink}
\fi
\fi

\usepackage{algorithm}
\usepackage[noend]{algpseudocode}
\usepackage{amssymb}
\usepackage{booktabs}
\usepackage{breakcites}
\usepackage[noabbrev]{cleveref}
\usepackage{graphicx}
\usepackage{multirow}
\usepackage{subcaption}
\usepackage{tikz}
  \usetikzlibrary{positioning, shadows, arrows, backgrounds, scopes}
\usepackage{url}
\usepackage{xcolor}
\usepackage{xspace}

\makeatletter
\def\blfootnote{\xdef\@thefnmark{}\@footnotetext}
\makeatother

\usepackage[sf={lining,proportional},tt={variable,lining,tabular},rm={lining,proportional}]{cfr-lm}
\newcommand\ttl[1]{\texttt{\textlg{#1}}}

\algnewcommand{\IfThenElse}[3]{\State \algorithmicif\ #1\ \algorithmicthen\ #2\ \algorithmicelse\ #3}
\algnewcommand{\IfThen}[2]{\State \algorithmicif\ #1\ \algorithmicthen\ #2}
\algnewcommand\OR{\textbf{or}\xspace}
\algnewcommand\AND{\textbf{and}\xspace}
\algnewcommand\NOT{\textbf{not}\xspace}
\algnewcommand\Break{\textbf{break}\xspace}
\algnewcommand\Continue{\textbf{continue}\xspace}

\definecolor{RedOrange}{cmyk}{0, 0.77,0.87,0}
\definecolor{RedOrange}{HTML}{FF4433}
\definecolor{Cerulean}{HTML}{007BA7}
\definecolor{Plum}{HTML}{DDA0DD}
\definecolor{OliveGreen}{HTML}{808000}

\providecommand{\keywords}[1]{\textit{\textbf{Keywords:}} #1}

\begin{document}

\ifreport
  \title{Revising Apetrei's bounding volume hierarchy construction algorithm to \protect\\allow stackless traversal}
  \date{\today}
  \reportnum{ORNL/TM-2024/3259}
  \division{Computational Sciences and Engineering Division}

\else
  \title{Revising Apetrei's bounding volume hierarchy construction algorithm to allow stackless traversal}
\fi

\ifjournal
  \author{Andrey~Prokopenko,
          and~Damien~Lebrun-Grandi\'e
\thanks{This manuscript has been authored by UT--Battelle, LLC, under Contract No.
DE-AC05-00OR22725 with the U.S. Department of Energy. The United States
Government retains and the publisher, by accepting the article for publication,
acknowledges that the United States Government retains a non-exclusive, paid-up,
irrevocable, world-wide license to publish or reproduce the published form of
this manuscript, or allow others to do so, for United States Government
purposes.}
}

\else\ifreport
  \author{Andrey Prokopenko \and Damien Lebrun-Grandi\'e}
\else
  \author{%
    A.~Prokopenko\thanks{Oak Ridge National Laboratory}\enskip\orcidlink{0000-0003-3616-5504},
    D.~Lebrun-Grandi\'e\footnotemark[1]\enskip\orcidlink{0000-0003-1952-7219}
  }
  \date{}
\fi
\fi

\ifreport
  \frontmatter
  \tableofcontents
  \mainmatter
\else
  \maketitle
\fi

\begin{abstract}
  Stackless traversal is a technique to speed up range queries by avoiding
  usage of a stack during the tree traversal. One way to achieve that is to
  transform a given binary tree to store a left child and a skip-connection
  (also called an escape index). In general, this operation requires an additional tree
  traversal during the tree construction. For some tree structures, however, it
  is possible to achieve the same result at a reduced cost. We propose one such
  algorithm for a GPU hierarchy construction algorithm proposed by Karras
  in~\cite{karras2012}. Furthermore, we show that our algorithm also works with
  the improved algorithm proposed by Apetrei in~\cite{apetrei2014}, despite a
  different ordering of the internal nodes. We achieve that by modifying
  Apetrei's algorithm to restore the original Karras' ordering of the internal
  nodes. Using the modified algorithm, we show how to construct a hierarchy
  suitable for a stackless traversal in a single bottom-up pass.
\end{abstract}

\ifjournal
  \begin{IEEEkeywords}
  bounding volume hierarchy, stackless traversal, GPU
  \end{IEEEkeywords}
\else
  \unless\ifreport
    \keywords{bounding volume hierarchy, stackless traversal, GPU}
  \fi
\fi

\unless\ifreport
\blfootnote {%
This manuscript has been authored by UT-Battelle, LLC, under contract
DE-AC05-00OR22725 with the U.S. Department of Energy. The United States
Government retains and the publisher, by accepting the article for publication,
acknowledges that the United States Government retains a nonexclusive, paid-up,
irrevocable, world-wide license to publish or reproduce the published form of
this manuscript, or allow others to do so, for United States Government
purposes.
}
\fi

\section{Introduction}
\label{s:introduction}

\ifjournal
  \IEEEPARstart{T}{ree}
\else
  Tree
\fi
structures, such as bounding volume hierarchy (BVH), octrees and kd-trees, are
used to accelerate the search for close geometric objects. Such trees are used
in many applications, including computer graphics (ray tracing, collision
detection), molecular dynamics, geographic information systems, cosmology, and
others.


The emergence of GPU accelerators spurred efforts to develop highly parallel
versions of the tree algorithms. Reducing thread execution divergence
(executing different code) and data divergence (reading or writing disparate
locations in memory) is highly desirable in parallel implementations,
particularly for accelerators with thousands of threads (such as GPUs). The
construction phase, in particular, is especially challenging on GPUs. An idea
of parallelization of BVH construction by using a space-filling curve (called
linear BVH, or LBVH) was first proposed in~\cite{lauterbach2009}, with further
improvements in~\cite{pantaleoni2010,garanzha2011}. The first fully parallel
algorithm allowing construction of all internal nodes concurrently was
introduced in~\cite{karras2012}, and further improved in~\cite{apetrei2014}.
The latter algorithm is considered to be the fastest BVH construction
algorithm on GPUs. Both Karras' and Apetrei's algorithms are widely
used~\cite{howard2019,arborx2020}, and may serve as an intermediate step
for constructing a higher quality BVH~\cite{karras2013,domingues2015}.

Search indexes have to support different types of search queries. The
\emph{range} search finds all objects that intersect with a query object.
Examples of the range search include finding all objects within a certain
distance and finding all triangles in the scene that a ray intersects with.

Range search is typically implemented using a stack to keep track of the nodes
to traverse. However, usage of stacks is undesirable on GPUs as it may lead to
lower occupancy due to higher memory demands per thread. To avoid it,
researchers developed stackless traversal, a technique to avoid explicitly
managing a stack of node pointers for each thread during the traversal. The
approach in~\cite{torres2009} introduced an idea of a \emph{skip connection}
(also called \emph{escape index}), which is a node index where the traversal
should proceed if the intersection test with the current node is not satisfied,
or if the node is a leaf node.

\begin{figure}[t]
  \centering
  \ifreport
    \resizebox{0.7\textwidth}{!}{%
      \input{figures/karras_ropes}
    }
  \else
    \resizebox{0.5\textwidth}{!}{%
      \input{figures/karras_ropes}
    }
  \fi
  \caption{An example of a hierarchy with skip-connections. The right children
  (dotted lines) are replaced with skip-connections (blue curves). The
  skip-connections on the right-most path point to the sentinel node
  $S$.}\label{f:karras_ropes}
\end{figure}

Stackless traversal requires a modification of a hierarchy, replacing right
children with skip connections. Typically, this requires an extra tree
traversal pass during the construction. In this work, we show that Karras'
internal node numbering allows a short calculation of the skip-connections.

While \cite{karras2012} and~\cite{apetrei2014} result in an identical hierarchy
structure, they produce different ordering of the internal nodes. Compared to
the Karras' ordering, in which siblings have subsequent indices, the siblings in
the Apetrei ordering may have an arbitrary large gap between them. The
unfortunate side effect of this new ordering is that the skip-connections can
no longer be easily set.

In the paper, we propose an elegant modification to the Apetrei's algorithm which
restores the original Karras node ordering. The proposed fix requires only a
slight algorithm change. It retains the performance gains in the hierarchy
construction, and allows construction of a hierarchy with the skip connections
in a single bottom-up pass.

To summarize, our key contributions are:
\begin{itemize}
  \item We demonstrate a straightforward way to determine skip-connections in Karras'
    algorithm.
  \item We show how Apetrei's algorithm can be modified to restore Karras'
    internal node ordering.
  \item We provide a single bottom-up traversal algorithm to generate a
    hierarchy with skip-connections.
  \item We include pseudo-codes for both construction and traversal algorithms
    to allow readers to easily implement them in their code.
\end{itemize}

The remainder of the paper is organized as follows. \Cref{s:stackless} provides
an overview of the stackless traversal. In \Cref{s:ordering}, we provide an
overview of both Karras' and Apetrei's algorithms and highlight their
differences. We describe our modification to the Apetrei's algorithm and
provide the resulting single bottom-up construction method in \Cref{s:apetrei_mod}.
Finally, we mention our implementation in \Cref{s:implementation}.

\section{Stackless traversal}
\label{s:stackless}

Stackless traversal is a technique to avoid explicitly managing a stack of node
pointers for each thread. The approach in~\cite{torres2009} introduced
\emph{skip-connections} (also called \emph{escape index}), an index of a node
where the traversal should proceed if the intersection test with the current
node is not satisfied, or if the node is a leaf node. In \Cref{f:karras_ropes},
the right children of the internal nodes are removed (denoted by dotted lines),
and skip-connections (blue curves) are introduced. For the nodes on the right-most path,
the skip-connections point to the artificial terminal node called \emph{sentinel}.

A critical observation is that each skip-connections points to the right child
of the last internal node that a given node is in the left subtree of. The
only exceptions to this rule are the nodes on the right-most path (including
the root node), which all point to the sentinel.

\input{algorithms/stackless_traversal}

\Cref{a:traversal} demonstrates the stackless traversal for a range search. If
an encountered node does not satisfy the predicate, its subtree is avoided by
immediately using skip-connection. Otherwise, either the left child is explored
next (for internal nodes), or a positive match is processed (leaf node).

\section{A tale of two orderings}
\label{s:ordering}

\begin{figure*}[t]
  \centering
  \begin{subfigure}[t]{0.45\textwidth}
    \centering
    \resizebox{\textwidth}{!}{%
      \input{figures/karras}
    }
    \caption{Karras\label{f:ordering:karras}}
  \end{subfigure}
  \qquad
  \begin{subfigure}[t]{0.45\textwidth}
    \centering
    \resizebox{\textwidth}{!}{
      \input{figures/apetrei}
    }
    \caption{Apetrei\label{f:ordering:apetrei}}
  \end{subfigure}
  \caption{The ordering of the internal nodes in Karras' and Apetrei's algorithms.}
\end{figure*}

\subsection{Karras' algorithm}
\label{s:karras}

Given $n$ primitives, the construction algorithm proposed in~\cite{karras2012}
is done in several steps:
\begin{enumerate}
  \setlength{\itemsep}{0pt}
  \item calculate Morton indices for the primitives;
  \item sort Morton indices;
  \item generate hierarchy structure;
  \item compute the bounding boxes of the internal nodes.
\end{enumerate}
The first two steps produce sorted Morton indices $\mathcal{M} =
\{m_i\}_{i=0}^{n-1}$ of the provided geometric objects. In step 3, the
algorithm constructs a binary radix tree as a hierarchical representation of
the common prefixes of a given set of keys (Morton indices in this case). The
constructed hierarchy has $n$ leaf and $n-1$ internal nodes.

Let $\delta(i, j)$ function be the longest common prefix between keys $m_i$ and
$m_j$ for $0 \le i < j < n$, and $+\infty$ for all other indices. We also
define $\delta^{*}(i) = \delta(i, i+1)$ for convenience.

The Karras' idea is that each internal node covers a linear range of keys, and
partitions its keys according to the highest differing bit in its range. The
split position $\gamma$ for an internal node covering the range $[i,j]$, $0 \le
i < j < n$, must satisfy $\delta(\gamma, \gamma+1) = \delta(i,j)$. The ranges
of the children of this internal node are then $[i, \gamma]$ and $[\gamma+1,
j]$.

To perform step 3 completely in parallel, the algorithm assigns internal
node indices to correspond to the split position in their parent. The children
of an internal node with a split $\gamma$ are assigned indices $\gamma$ and
$\gamma+1$ in either the internal node array $\mathcal{I} =
\left\{I_k\right\}_{k=0}^{n-2}$, or the leaf node array $\mathcal{L} =
\left\{L_k\right\}_{k=0}^{n-1}$, assuming they are stored separately. This way,
an internal node information can be fully ascertained by its range and its
split. Karras' algorithm determines these values using linear and binary search
through $\mathcal{M}$ using $\delta$ function. The root node is
assigned index 0. For more details, see~\cite{karras2012}.

In the Karras layout, the index of each internal
node coincides with one of the bounds of its range. Specifically, if an
internal node with a range $[i,j]$ is a left child of its parent, its index is
$j$; otherwise, if it is a right child, it is $i$. It can also be seen that it
coincides with the range bound that has a smaller out of two values
$\delta^*(i-1)$ and $\delta^*(j)$ (the root node index is always that of its
left range bound). We will use this property to modify the Apetrei's algorithm.

An example of a constructed hierarchy for a set of Morton indices is shown
in~\Cref{f:ordering:karras} ($n = 8$). The internal nodes $\mathcal{I}$ are
shown in orange, leaf nodes $\mathcal{L}$ are in green, and the split position
for each internal node in red. The ranges for the internal nodes are denoted by
gray boxes. For instance, internal node $I_3$ covers the range $[0, 3]$ and has
the split position $\gamma = 1$, thus having two children $I_1$ and $I_2$ (both
internal nodes). Similarly, $I_4$ covers the range $[4, 7]$ with a split
$\gamma = 4$ and children $L_4$ and $I_5$ (one leaf and one internal node).

\input{algorithms/karras_ropes}

One can observe, that for the Karras' node ordering, the index of the target
node of a skip connection is going to be the right range of the nodes spanned
by its parent, incremented by one. \Cref{a:karras_ropes} shows the calculation
of the skip-connection target for a given node $N$. If the right boundary range
corresponds to the rightmost leaf, this indicates that the node is on the
right-most side, so that its skip-connection should point to the sentinel node.
For all other nodes, the skip-connection will point to the node whose index the
right range boundary incremented by one. While the index is straightforward to determine,
an additional calculation (\cref{l:karras_ropes:is_leaf}) is required to figure
out whether it belongs to a leaf or an internal node.

\input{algorithms/apetrei}

\subsection{Apetrei's algorithm}
\label{s:apetrei}

The algorithm modification proposed in~\cite{apetrei2014} merged steps 3 and 4 of the
Karras' algorithm into a single bottom-up step. In contrast with Karras
algorithm, where each internal node constructs its range and determines a split
independently, in Apetrei's approach the ranges get merged in the bottom-up
traversal starting from the leaves, resulting in a faster algorithm, while
also being easier to implement and requiring fewer lines of code. The internal
nodes are indexed using the split positions $\gamma$ rather than one of the
ends of the corresponding range. An interesting side effect of such ordering is
that the root node is no longer guaranteed to be $I_0$, and its index now needs
to be stored in the hierarchy.

An example of a resulting hierarchy and internal node ordering is shown
on~\Cref{f:ordering:apetrei}. As the algorithm uses splits for internal node
indices, the gap between indices of two siblings may become large. For example,
the children of $I_0$ have a gap of 3.

The pseudocode for the Apetrei's algorithm is shown in~\Cref{a:apetrei}.
Compared to the original paper, the presented version shows the complete
procedure, except for the construction of the bounding boxes.
The $\textsc{atomicCAS}$ function performs the atomic Compare-And-Swap
operation. It compares the contents of a memory location (first argument) with
a given value (second argument). If they are the same, it overwrites the
contents of that location with a new given value (third argument). This is done
as a single atomic operation. It returns the value read from the memory
location (\textit{not} the value written to it). By comparing the return value
of $\textsc{atomicCAS}$ with the initialization value of the $\ttl{store}$
array, it can be determined whether that location has already been modified. If
it was, it indicates that the current thread is the second thread up and may
proceed further. Otherwise, as the first thread up, the thread exits the procedure.

The~\Cref{a:apetrei} includes several additional optimizations not present
in~\cite{apetrei2014}. First, the algorithm keeps track of the results of
$\delta^*$ for the ranges, updating them only when necessary. This results in
fewer memory loads of the Morton indices array which exhibit a random access
pattern\footnote{We found that computing $\delta^*$ values as part of the
bottom-up procedure is faster than pre-computing and storing them beforehand.}.
Second, the temporary storage serves dual purpose, both as a flag for allowing
only one thread up, as well as for temporary storage of the opposite range,
reducing memory allocation. We also note that, as recommended
in~\cite{apetrei2014}, the $\delta^*$ function is switched from computing the
common prefix (as in the Karras' algorithm) to a simpler XOR evaluation. If the
Morton codes are identical, we follow the Karras idea of augmenting the key
with a bit representation of its index.

\unless\ifreport
\input{algorithms/apetrei_mod_with_ropes}
\fi

\section{Modified Apetrei's algorithm with skip-connections}
\label{s:apetrei_mod}

\ifreport
\input{algorithms/apetrei_mod_with_ropes}
\fi

We modify the Apetrei's algorithm to restore the Karras ordering of the
internal nodes and installation of skip-connections. The modified version is
presented in~\Cref{a:apetrei_mod_with_ropes}. Let us highlight the differences
with \Cref{a:apetrei}.

First, while the Apetrei index $p$ of the parent node is still being calculated
and used for referencing the temporary storage, the Karras index $q$
is now used to reference the location of the parent node. The index computation
on~\cref{l:apetrei_mod:karras} of~\Cref{a:apetrei_mod_with_ropes} uses the property we
mentioned in~\Cref{s:karras}. Specifically, that the Karras index coincides
with its range boundary that has the smaller value of $\delta^*$. Thus, by
comparing the $\delta^*$ values of the parent range, we are able to figure out
its Karras index.

Second, a parent node is now updated by a single thread instead of two.
Specifically, \cref{l:apetrei:left_child,l:apetrei:right_child}
in~\Cref{a:apetrei} are replaced by
lines~\ref{l:apetrei_mod:parent_update_begin}--\ref{l:apetrei_mod:parent_update_end}
in~\Cref{a:apetrei_mod_with_ropes}. This is due to the fact that the
determination of the parent index, $q$, now requires the knowledge of its full
range, which is only available to the second thread up. On the flip side, a
parent node can now determine the indices of its children, as knowing a split
position $\gamma$ (which is exactly the Apetrei index $p$ in this case), the
indices of the children are $\gamma$ and $\gamma+1$ in the internal or the leaf
node array.

Third, given the knowledge of the Karras index $q$ and the full range of the
parent as part of the Apetrei's algorithm, we can now easily calculate the
target for each of the skip-connections as was explained in \Cref{s:karras}.
Lines~\ref{l:apetrei_mod:leaf_update_begin}--\ref{l:apetrei_mod:leaf_update_end} (lines \ref{l:apetrei_mod:parent_rope_update_begin}--\ref{l:apetrei_mod:parent_update_end}) integrate \Cref{a:karras_ropes} into the procedure for leaf (internal) nodes, respectively.

Finally, the index $i$ on~\cref{l:apetrei_mod:i_update} is updated with the
Karras' index instead of Apetrei's, and the loop termination condition is
simplified, as $I_0$ is always the root in the Karras ordering.

\ifexperiments
\section{Performance results}

In this Section, we present the numerical results comparing the performance of
the algorithms on several problems.

In our experiments, we used several of the experimental data sets
from~\cite{arborx2020}, a filled and hollow boxes. To generate $n$ points for
these point clouds, set $a = p^{1/3}$, $\Omega = [-a,a]^3$, and proceed as
follows:
\begin{itemize}
  \setlength{\itemsep}{0pt}
  \item
    \textit{filled cube:} each random point is drawn randomly from $\Omega$ with uniform
    distribution;
  \item
    \textit{hollow cube:} points are placed on the faces of $\Omega$ in a cyclic manner,
    with the position of the point on each face being random with uniform
    distribution;
\end{itemize}

The described algorithms were implemented in the ArborX
library~\cite{arborx2020}. The experiments were performed on a single node of
the OLCF Summit system using a single Nvidia Volta V100 GPU~\cite{olcf_summit}.
Google Benchmark tool~\cite{google_benchmark} was used for our experiments,
using the median of the 10 benchmark repetitions for the results we have
reported here.

\begin{figure}[t]
  \begin{center}
    \begin{subfigure}{0.45\textwidth}
      \includegraphics[width=\textwidth]{figures/summit_cuda_construction_filled_box.png}
      \caption{Filled case}
    \end{subfigure}
    \begin{subfigure}{0.45\textwidth}
      \includegraphics[width=\textwidth]{figures/summit_cuda_construction_hollow_box.png}
      \caption{Hollow case}
    \end{subfigure}
  \end{center}
  \caption{Comparison of the construction algorithms on V100 for the filled and hollow point clouds.}
  \label{f:construction}
\end{figure}

\Cref{f:construction} shows the construction rates (points/second) for a range
of $n$ values and two geometries. We confirmed that As stated in the
original~\cite{apetrei2014} paper, the Apetrei's algorithm "is strictly faster".
More surprisingly, the modified version of the Apetrei's algorithm outperforms
the original version. While it is not fully clear why that would be the case, a
reasonable assumption would be less random memory access due to the Karras node
ordering, as well as the fact that each node is updated by a single thread.
\fi

\section{Implementation}\label{s:implementation}

The described algorithm is implemented as part of the ArborX
library~\cite{arborx2020}. The code is available at
\url{https://github.com/arborx/ArborX}.

\unless\ifjournal
\unless\ifreport
\section*{CRediT author statement}

\textbf{Andrey Prokopenko}: Conceptualization, Investigation, Software, Writing - original draft.
\textbf{Damien Lebrun-Grandi\'e}: Software, Writing - review and editing.
\fi
\fi

\section*{Acknowledgements}
This research was supported by the Exascale Computing Project (17-SC-20-SC), a
collaborative effort of the U.S. Department of Energy Office of Science and
the National Nuclear Security Administration.

\ifexperiments
This research used resources of the Oak Ridge Leadership Computing Facility at
the Oak Ridge National Laboratory, which is supported by the Office of Science
of the U.S. Department of Energy under Contract No. DE-AC05-00OR22725.
\fi

\ifreport
  \section{REFERENCES}
  \printbibliography[heading=none]
  \backmatter
\else
  \ifjournal
    \bibliographystyle{IEEEtran}
  \else
    \bibliographystyle{apalike}
  \fi
\bibliography{main}
\fi

\end{document}

%% file: figures/karras_ropes.tex
\definecolor{ColorInternalIndex}{HTML}{E4C6B3}
\definecolor{ColorLeafIndex}{HTML}{C0C7B8}
\definecolor{ColorSplit}{HTML}{D48C8B}
\definecolor{ColorNode}{HTML}{ECECEC}
\definecolor{ColorGray}{HTML}{666666}
\definecolor{ColorSentinel}{HTML}{DBDB8D}
\definecolor{ColorRope}{HTML}{AEC7E8}

\begin{tikzpicture}[
    box/.style={rectangle, draw=black, fill=ColorNode, text=ColorNode, text centered, minimum height=0.3cm},
    v_box/.style={rectangle, draw=black, fill=ColorNode, text=ColorGray, align=center},
    internal_parent/.style={circle, draw=black, fill=ColorSplit, text=ColorSplit},
    internal_node/.style={circle, draw=black, fill=ColorInternalIndex, text=black},
    leaf_node/.style={circle, draw=black, fill=ColorLeafIndex, text=black},
    sentinel_node/.style={circle, draw=black, fill=ColorSentinel, text=black},
    text centered,
    font=\bf,
    anchor=north west,
    level distance=0.5cm,
    growth parent anchor=south,
    line width=1pt,
    >=stealth,
    tips=proper
  ]
  {[on background layer]
    \node(box_0)[box, minimum width=11.0cm, opacity=0] at (0.0, 0){};
    \node(box_3)[box, minimum width= 5.0cm, opacity=0] at (0.0,-2){};
    \node(box_4)[box, minimum width= 5.0cm, opacity=0] at (6.0,-1){};
    \node(box_1)[box, minimum width= 2.0cm, opacity=0] at (0.0,-3){};
    \node(box_2)[box, minimum width= 2.0cm, opacity=0] at (3.0,-3){};
    \node(box_5)[box, minimum width= 3.5cm, opacity=0] at (7.5,-2){};
    \node(box_6)[box, minimum width= 2.0cm, opacity=0] at (7.5,-3){};
  }
  \node(internal_node_0)[internal_node, left =-0.6cm of box_0, opacity=0] {0};
  \node(internal_node_3)[internal_node, right=-0.6cm of box_3, opacity=0] {3};
  \node(internal_node_4)[internal_node, left =-0.6cm of box_4, opacity=0] {4};
  \node(internal_node_1)[internal_node, right=-0.6cm of box_1, opacity=0] {1};
  \node(internal_node_2)[internal_node, left =-0.6cm of box_2, opacity=0] {2};
  \node(internal_node_5)[internal_node, left =-0.6cm of box_5, opacity=0] {5};
  \node(internal_node_6)[internal_node, right=-0.6cm of box_6, opacity=0] {6};
  \node(internal_parent_0)[internal_node, right=4.75cm of internal_node_0] {0};
  \node(internal_parent_3)[internal_node, left=1.75cm of internal_node_3] {3};
  \node(internal_parent_4)[internal_node, right=0.25cm of internal_node_4] {4};
  \node(internal_parent_1)[internal_node, left=0.25cm of internal_node_1] {1};
  \node(internal_parent_2)[internal_node, right=0.25cm of internal_node_2] {2};
  \node(internal_parent_5)[internal_node, right=1.75cm of internal_node_5] {5};
  \node(internal_parent_6)[internal_node, left=0.25cm of internal_node_6] {6};
  \node(leaf_node_0)[leaf_node] at ( 0.0,-4) {0};
  \node(leaf_node_1)[leaf_node] at ( 1.5,-4) {1};
  \node(leaf_node_2)[leaf_node] at ( 3.0,-4) {2};
  \node(leaf_node_3)[leaf_node] at ( 4.5, -4) {3};
  \node(leaf_node_4)[leaf_node] at ( 6.0, -4) {4};
  \node(leaf_node_5)[leaf_node] at ( 7.5, -4) {5};
  \node(leaf_node_6)[leaf_node] at ( 9.0, -4) {6};
  \node(leaf_node_7)[leaf_node] at (10.5, -4) {7};
  \node(sentinel)[sentinel_node] at ( 12.0,-1) {S};
  \draw [->] (internal_parent_0) edge           (internal_parent_3);
  \draw [->] (internal_parent_0) edge[dotted]   (internal_parent_4);
  \draw [->] (internal_parent_3) edge           (internal_parent_1);
  \draw [->] (internal_parent_3) edge[dotted]   (internal_parent_2);
  \draw [->] (internal_parent_4) edge           (leaf_node_4);
  \draw [->] (internal_parent_4) edge[dotted]   (internal_parent_5);
  \draw [->] (internal_parent_1) edge           (leaf_node_0);
  \draw [->] (internal_parent_1) edge[dotted]   (leaf_node_1);
  \draw [->] (internal_parent_2) edge           (leaf_node_2);
  \draw [->] (internal_parent_2) edge[dotted]   (leaf_node_3);
  \draw [->] (internal_parent_5) edge           (internal_parent_6);
  \draw [->] (internal_parent_5) edge[dotted]   (leaf_node_7);
  \draw [->] (internal_parent_6) edge           (leaf_node_5);
  \draw [->] (internal_parent_6) edge[dotted]   (leaf_node_6);
  \draw [->,ColorRope] (internal_parent_0) to[bend left]  (sentinel);
  \draw [->,ColorRope] (internal_parent_3) to[bend left]  (internal_parent_4);
  \draw [->,ColorRope] (internal_parent_4) to[bend left]  (sentinel);
  \draw [->,ColorRope] (internal_parent_1) to[bend left]  (internal_parent_2);
  \draw [->,ColorRope] (internal_parent_2) to[bend left]  (internal_parent_4);
  \draw [->,ColorRope] (internal_parent_5) to[bend left]  (sentinel);
  \draw [->,ColorRope] (internal_parent_6) to[bend left]  (leaf_node_7);
  \draw [->,ColorRope] (leaf_node_0)       to[bend left]  (leaf_node_1);
  \draw [->,ColorRope] (leaf_node_1)       to[bend left]  (internal_parent_2);
  \draw [->,ColorRope] (leaf_node_2)       to[bend left]  (leaf_node_3);
  \draw [->,ColorRope] (leaf_node_3)       to[bend left]  (internal_parent_4);
  \draw [->,ColorRope] (leaf_node_4)       to[bend left]  (internal_parent_5);
  \draw [->,ColorRope] (leaf_node_5)       to[bend left]  (leaf_node_6);
  \draw [->,ColorRope] (leaf_node_6)       to[bend left]  (leaf_node_7);
  \draw [->,ColorRope] (leaf_node_7)       to[bend left]  (sentinel);
\end{tikzpicture}

%% file: algorithms/stackless_traversal.tex
\begin{algorithm}[t]
  \caption{Stackless tree traversal algorithm using skip-connections for a
  predicate $query$. Each node $N$ stores a skip-connection $N_{skip}$. In
  addition, each internal node stores a left child $N_{left}$. $\blacklozenge$
  denotes the sentinel node.}\label{a:traversal}
  \begin{algorithmic}[1]
    \State $N \gets I_0$ \Comment{Start from the root node}
    \Repeat
      \If {$query$ is satisfied on $N$}
        \If {$N$ is a leaf node}
          \State {Store the result or perform an operation}
          \State {$N \gets N_{skip}$}
        \Else
          \State {$N \gets N_{left}$}
        \EndIf
      \Else
        \State {$N \gets N_{skip}$}
      \EndIf
    \Until{$N = \blacklozenge$}
  \end{algorithmic}
\end{algorithm}

%% file: figures/karras.tex
\definecolor{ColorInternalIndex}{HTML}{E4C6B3}
\definecolor{ColorLeafIndex}{HTML}{C0C7B8}
\definecolor{ColorSplit}{HTML}{D48C8B}
\definecolor{ColorNode}{HTML}{ECECEC}
\definecolor{ColorGray}{HTML}{666666}

\begin{tikzpicture}[
    box/.style={rectangle, draw=black, fill=ColorNode, text=ColorNode, text centered, minimum height=0.3cm},
    v_box/.style={rectangle, draw=black, fill=ColorNode, text=ColorGray, align=center},
    internal_parent/.style={circle, draw=black, fill=ColorSplit, text=ColorSplit},
    internal_node/.style={circle, draw=black, fill=ColorInternalIndex, text=black},
    leaf_node/.style={circle, draw=black, fill=ColorLeafIndex, text=black},
    text centered,
    font=\bf,
    anchor=north west,
    level distance=0.5cm,
    growth parent anchor=south,
    line width=1pt,
    >=stealth,
    tips=proper
  ]
  {[on background layer]
    \node(box_0)[box, minimum width=11.0cm] at (0.0, 0){};
    \node(box_3)[box, minimum width= 5.0cm] at (0.0,-2){};
    \node(box_4)[box, minimum width= 5.0cm] at (6.0,-1){};
    \node(box_1)[box, minimum width= 2.0cm] at (0.0,-3){};
    \node(box_2)[box, minimum width= 2.0cm] at (3.0,-3){};
    \node(box_5)[box, minimum width= 3.5cm] at (7.5,-2){};
    \node(box_6)[box, minimum width= 2.0cm] at (7.5,-3){};
  }
  \node(internal_node_0)[internal_node, left =-0.6cm of box_0] {0};
  \node(internal_node_3)[internal_node, right=-0.6cm of box_3] {3};
  \node(internal_node_4)[internal_node, left =-0.6cm of box_4] {4};
  \node(internal_node_1)[internal_node, right=-0.6cm of box_1] {1};
  \node(internal_node_2)[internal_node, left =-0.6cm of box_2] {2};
  \node(internal_node_5)[internal_node, left =-0.6cm of box_5] {5};
  \node(internal_node_6)[internal_node, right=-0.6cm of box_6] {6};
  \node(internal_parent_0)[internal_parent, right=4.75cm of internal_node_0] {};
  \node(internal_parent_3)[internal_parent, left=1.75cm of internal_node_3] {};
  \node(internal_parent_4)[internal_parent, right=0.25cm of internal_node_4] {};
  \node(internal_parent_1)[internal_parent, left=0.25cm of internal_node_1] {};
  \node(internal_parent_2)[internal_parent, right=0.25cm of internal_node_2] {};
  \node(internal_parent_5)[internal_parent, right=1.75cm of internal_node_5] {};
  \node(internal_parent_6)[internal_parent, left=0.25cm of internal_node_6] {};
  \node(leaf_node_0)[leaf_node] at ( 0.0,-4) {0};
  \node(leaf_node_1)[leaf_node] at ( 1.5,-4) {1};
  \node(leaf_node_2)[leaf_node] at ( 3.0,-4) {2};
  \node(leaf_node_3)[leaf_node] at ( 4.5, -4) {3};
  \node(leaf_node_4)[leaf_node] at ( 6.0, -4) {4};
  \node(leaf_node_5)[leaf_node] at ( 7.5, -4) {5};
  \node(leaf_node_6)[leaf_node] at ( 9.0, -4) {6};
  \node(leaf_node_7)[leaf_node] at (10.5, -4) {7};
  \draw [->] (internal_parent_0) edge (internal_node_3);
  \draw [->] (internal_parent_0) edge (internal_node_4);
  \draw [->] (internal_parent_3) edge (internal_node_1);
  \draw [->] (internal_parent_3) edge (internal_node_2);
  \draw [->] (internal_parent_4) edge (leaf_node_4);
  \draw [->] (internal_parent_4) edge (internal_node_5);
  \draw [->] (internal_parent_1) edge (leaf_node_0);
  \draw [->] (internal_parent_1) edge (leaf_node_1);
  \draw [->] (internal_parent_2) edge (leaf_node_2);
  \draw [->] (internal_parent_2) edge (leaf_node_3);
  \draw [->] (internal_parent_5) edge (internal_node_6);
  \draw [->] (internal_parent_5) edge (leaf_node_7);
  \draw [->] (internal_parent_6) edge (leaf_node_5);
  \draw [->] (internal_parent_6) edge (leaf_node_6);

  {[on background layer]
    \node (v_box_0)[v_box] at (0.0,  -4.2) {\,\\\,\\0\\0\\0\\0\\1};
    \node (v_box_1)[v_box] at (1.5,  -4.2) {\,\\\,\\0\\0\\0\\1\\0};
    \node (v_box_2)[v_box] at (3.0,  -4.2) {\,\\\,\\0\\0\\1\\0\\0};
    \node (v_box_3)[v_box] at (4.5,  -4.2) {\,\\\,\\0\\0\\1\\0\\1};
    \node (v_box_4)[v_box] at (6.0,  -4.2) {\,\\\,\\1\\0\\0\\1\\1};
    \node (v_box_5)[v_box] at (7.5,  -4.2) {\,\\\,\\1\\1\\0\\0\\0};
    \node (v_box_6)[v_box] at (9.0,  -4.2) {\,\\\,\\1\\1\\0\\0\\1};
    \node (v_box_7)[v_box] at (10.5, -4.2) {\,\\\,\\1\\1\\1\\1\\0};
  }

  \ifreport
    \draw[<->] (0.5,-6.60) edge (1.45,-6.60);
    \draw[<->] (2.0,-6.15) edge (2.95,-6.15);
    \draw[<->] (3.5,-7.10) edge (4.45,-7.10);
    \draw[<->] (5.0,-5.15) edge (5.95,-5.15);
    \draw[<->] (6.5,-5.65) edge (7.45,-5.65);
    \draw[<->] (8.0,-7.10) edge (8.95,-7.10);
    \draw[<->] (9.5,-6.15) edge (10.45,-6.15);
  \else
    \draw[<->] (0.5,-6.35) edge (1.45,-6.35);
    \draw[<->] (2.0,-5.90) edge (2.95,-5.90);
    \draw[<->] (3.5,-6.75) edge (4.45,-6.75);
    \draw[<->] (5.0,-5.10) edge (5.95,-5.1);
    \draw[<->] (6.5,-5.50) edge (7.45,-5.5);
    \draw[<->] (8.0,-6.75) edge (8.95,-6.75);
    \draw[<->] (9.5,-5.95) edge (10.45,-5.95);
  \fi
\end{tikzpicture}

%% file: figures/apetrei.tex
\definecolor{ColorInternalIndex}{HTML}{E4C6B3}
\definecolor{ColorLeafIndex}{HTML}{C0C7B8}
\definecolor{ColorSplit}{HTML}{D48C8B}
\definecolor{ColorNode}{HTML}{ECECEC}
\definecolor{ColorGray}{HTML}{666666}

\begin{tikzpicture}[
    box/.style={rectangle, draw=black, fill=ColorNode, text=ColorNode, text centered, minimum height=0.3cm},
    v_box/.style={rectangle, draw=black, fill=ColorNode, text=ColorGray, align=center},
    internal_parent/.style={circle, draw=black, fill=ColorSplit, text=black},
    internal_node/.style={circle, draw=black, fill=ColorInternalIndex, text=black},
    leaf_node/.style={circle, draw=black, fill=ColorLeafIndex, text=black},
    text centered,
    font=\bf,
    anchor=north west,
    level distance=0.5cm,
    growth parent anchor=south,
    line width=1pt,
    >=stealth,
    tips=proper
  ]
  {[on background layer]
    \node(box_0)[box, minimum width=11.0cm] at (0.0, 0){};
    \node(box_3)[box, minimum width= 5.0cm] at (0.0,-2){};
    \node(box_4)[box, minimum width= 5.0cm] at (6.0,-1){};
    \node(box_1)[box, minimum width= 2.0cm] at (0.0,-3){};
    \node(box_2)[box, minimum width= 2.0cm] at (3.0,-3){};
    \node(box_5)[box, minimum width= 3.5cm] at (7.5,-2){};
    \node(box_6)[box, minimum width= 2.0cm] at (7.5,-3){};
  }
  \node(internal_node_0)[internal_node, left =-0.3cm of box_0] {};
  \node(internal_node_3)[internal_node, right=-0.3cm of box_3] {};
  \node(internal_node_4)[internal_node, left =-0.3cm of box_4] {};
  \node(internal_node_1)[internal_node, right=-0.3cm of box_1] {};
  \node(internal_node_2)[internal_node, left =-0.3cm of box_2] {};
  \node(internal_node_5)[internal_node, left =-0.3cm of box_5] {};
  \node(internal_node_6)[internal_node, right=-0.3cm of box_6] {};
  \node(leaf_node_0)[leaf_node] at ( 0.0,-4) {0};
  \node(leaf_node_1)[leaf_node] at ( 1.5,-4) {1};
  \node(leaf_node_2)[leaf_node] at ( 3.0,-4) {2};
  \node(leaf_node_3)[leaf_node] at ( 4.5, -4) {3};
  \node(leaf_node_4)[leaf_node] at ( 6.0, -4) {4};
  \node(leaf_node_5)[leaf_node] at ( 7.5, -4) {5};
  \node(leaf_node_6)[leaf_node] at ( 9.0, -4) {6};
  \node(leaf_node_7)[leaf_node] at (10.5, -4) {7};
  \node(internal_parent_0)[internal_parent, left=0.4cm of internal_node_1] {0};
    \draw [->] (internal_parent_0) edge (leaf_node_0);
    \draw [->] (internal_parent_0) edge (leaf_node_1);
  \node(internal_parent_2)[internal_parent, right=0.4cm of internal_node_2] {2};
    \draw [->] (internal_parent_2) edge (leaf_node_2);
    \draw [->] (internal_parent_2) edge (leaf_node_3);
  \node(internal_parent_5)[internal_parent, left=0.4cm of internal_node_6] {5};
    \draw [->] (internal_parent_5) edge (leaf_node_5);
    \draw [->] (internal_parent_5) edge (leaf_node_6);
  \node(internal_parent_1)[internal_parent, left=1.9cm of internal_node_3] {1};
    \draw [->] (internal_parent_1) edge (internal_parent_0);
    \draw [->] (internal_parent_1) edge (internal_parent_2);
  \node(internal_parent_6)[internal_parent, right=1.9cm of internal_node_5] {6};
    \draw [->] (internal_parent_6) edge (internal_parent_5);
    \draw [->] (internal_parent_6) edge (leaf_node_7);
  \node(internal_parent_4)[internal_parent, right=0.2cm of internal_node_4] {4};
    \draw [->] (internal_parent_4) edge (internal_parent_6);
    \draw [->] (internal_parent_4) edge (leaf_node_4);
  \node(internal_parent_3)[internal_parent, right=4.9cm of internal_node_0] {3};
    \draw [->] (internal_parent_3) edge (internal_parent_1);
    \draw [->] (internal_parent_3) edge (internal_parent_4);

  {[on background layer]
    \node (v_box_0)[v_box] at (0.0, -4.25)  {\,\\\,\\0\\0\\0\\0\\1};
    \node (v_box_1)[v_box] at (1.5, -4.25)  {\,\\\,\\0\\0\\0\\1\\0};
    \node (v_box_2)[v_box] at (3.0, -4.25)  {\,\\\,\\0\\0\\1\\0\\0};
    \node (v_box_3)[v_box] at (4.5, -4.25)  {\,\\\,\\0\\0\\1\\0\\1};
    \node (v_box_4)[v_box] at (6.0, -4.25)  {\,\\\,\\1\\0\\0\\1\\1};
    \node (v_box_5)[v_box] at (7.5, -4.25)  {\,\\\,\\1\\1\\0\\0\\0};
    \node (v_box_6)[v_box] at (9.0, -4.25)  {\,\\\,\\1\\1\\0\\0\\1};
    \node (v_box_7)[v_box] at (10.5, -4.25) {\,\\\,\\1\\1\\1\\1\\0};
  }

  \ifreport
    \draw[<->] (0.5,-6.60) edge (1.45,-6.60);
    \draw[<->] (2.0,-6.15) edge (2.95,-6.15);
    \draw[<->] (3.5,-7.10) edge (4.45,-7.10);
    \draw[<->] (5.0,-5.15) edge (5.95,-5.15);
    \draw[<->] (6.5,-5.65) edge (7.45,-5.65);
    \draw[<->] (8.0,-7.10) edge (8.95,-7.10);
    \draw[<->] (9.5,-6.15) edge (10.45,-6.15);
  \else
    \draw[<->] (0.5,-6.35) edge (1.45,-6.35);
    \draw[<->] (2.0,-5.95) edge (2.95,-5.95);
    \draw[<->] (3.5,-6.75) edge (4.45,-6.75);
    \draw[<->] (5.0,-5.10) edge (5.95,-5.1);
    \draw[<->] (6.5,-5.50) edge (7.45,-5.5);
    \draw[<->] (8.0,-6.75) edge (8.95,-6.75);
    \draw[<->] (9.5,-5.95) edge (10.45,-5.95);
  \fi
\end{tikzpicture}

%% file: algorithms/karras_ropes.tex
\begin{algorithm}[t]
  \caption{Skip-connections in Karras' node ordering for a node $N$.
  $\blacklozenge$ denotes the sentinel node.}\label{a:karras_ropes}
  \begin{algorithmic}[1]
     \If{$\ttl{range}_{right} = n-1$}
       \State $N_{skip} \gets \blacklozenge$
     \Else
       \State {$r \gets \ttl{range}_{right} + 1$}
       \If{$\delta^*(r-1) < \delta^*(r)$} \label{l:karras_ropes:is_leaf}
          \State {$N_{skip} \gets L_r$}
       \Else
          \State {$N_{skip} \gets I_r$}
       \EndIf
      \EndIf
  \end{algorithmic}
\end{algorithm}

%% file: algorithms/apetrei.tex
\begin{algorithm*}[t]
  \caption{Apetrei's algorithm. For simplicity, the construction of the
  bounding boxes is omitted. $\diamondsuit$ denotes an invalid entry value
  (e.g., $-1$).}\label{a:apetrei}
  \begin{algorithmic}[1]
    \State Initialize all entries in $\ttl{store}$ to $\diamondsuit$
    \ForAll {leaf node with index $i \in [0, n-1]$ \textbf{in parallel}}
      \State $\ttl{range}_{left} \gets i$, \quad\, $\delta_{left} \gets \delta^*(\ttl{range}_{left}-1)$
      \State $\ttl{range}_{right} \gets i$, \quad $\delta_{right} \gets \delta^*(\ttl{range}_{right})$
      \Repeat
        \If {$\delta_{right} < \delta_{left}$}                  \Comment Left child of its parent
          \State $p \gets \ttl{range}_{right}$                  \Comment Apetrei index $p$
          \IfThenElse{$\ttl{range}_{left} = \ttl{range}_{right}$}
                     {$I_{p,left} \gets L_i$}
                     {$I_{p,left} \gets I_i$}                   \label{l:apetrei:left_child}
          \State $\ttl{range}_{right} \gets \textsc{atomicCAS}(\ttl{store}_p, \diamondsuit, \ttl{range}_{left})$
          \IfThen{$\ttl{range}_{right} = \diamondsuit$}{\Return}
          \State $\delta_{right} \gets \delta^*(\ttl{range}_{right})$ \Comment Recompute outdated value
        \Else                                                   \Comment Right child of its parent $p$
          \State $p \gets \ttl{range}_{left}-1$
          \IfThenElse{$\ttl{range}_{left} = \ttl{range}_{right}$}
                     {$I_{p,right} \gets L_i$}
                     {$I_{p,right} \gets I_i$}                   \label{l:apetrei:right_child}
          \State $\ttl{range}_{left} \gets \textsc{atomicCAS}(\ttl{store}_p, \diamondsuit, \ttl{range}_{right})$
          \IfThen{$\ttl{range}_{left} = \diamondsuit$}{\Return}
          \State $\delta_{left} \gets \delta^*(\ttl{range}_{left}-1)$ \Comment Recompute outdated value
        \EndIf
        \State $i \gets p$
      \Until{$r_i^{right} = r_i^{left} + n-1$}
    \EndFor
  \end{algorithmic}
\end{algorithm*}

%% file: algorithms/apetrei_mod_with_ropes.tex
\begin{algorithm*}[t]
  \caption{Modified Apetrei's algorithm with skip installation. For simplicity,
  the construction of the bounding boxes is omitted. $\diamondsuit$ denotes an
  invalid entry value (e.g., $-1$). $\blacklozenge$ denotes the sentinel
  node.}\label{a:apetrei_mod_with_ropes}
  \begin{algorithmic}[1]
    \State Initialize all entries in $\ttl{store}$ to $\diamondsuit$
    \ForAll {leaf node with index $i \in [0, n-1]$ \textbf{in parallel}}
      \State $\ttl{range}_{left} \gets i$, \quad\, $\delta_{left} \gets \delta^*(i-1)$
      \State $\ttl{range}_{right} \gets i$, \quad $\delta_{right} \gets \delta^*(i)$
      \If{$i = n-1$}\label{l:apetrei_mod:leaf_update_begin}
        \State $L_{i,skip} \gets \blacklozenge$
      \Else
        \IfThenElse{$\delta_{right} < \delta^*(i+1)$}
                   {$L_{i,skip} \gets L_{i+1}$}
                   {$L_{i,skip} \gets I_{i+1}$}             \label{l:apetrei_mod:leaf_skip}
      \EndIf\label{l:apetrei_mod:leaf_update_end}
      \Repeat
        \If {$\delta_{right} < \delta_{left}$}                  \Comment Left child of its parent
          \State $p \gets \ttl{range}_{right}$                  \Comment Apetrei index $p$
          \State $\ttl{range}_{right} \gets \textsc{atomicCAS}(\ttl{store}_p, \diamondsuit, \ttl{range}_{left})$
          \IfThen{$\ttl{range}_{right} = \diamondsuit$}{\Return}
          \State $\delta_{right} \gets \delta^*(\ttl{range}_{right})$ \Comment Recompute outdated value
        \Else                                                   \Comment Right child of its parent
          \State $p \gets \ttl{range}_{left}-1$
          \State $\ttl{range}_{left} \gets \textsc{atomicCAS}(\ttl{store}_p, \diamondsuit, \ttl{range}_{right})$
          \IfThen{$\ttl{range}_{left} = \diamondsuit$}{\Return}
          \State $\delta_{left} \gets \delta^*(\ttl{range}_{left}-1)$ \Comment Recompute outdated value
        \EndIf
        \IfThenElse{$\delta_{right} < \delta_{left}$}
                   {$q \gets \ttl{range}_{right}$}
                   {$q \gets \ttl{range}_{left}$}      \Comment Karras index $q$\label{l:apetrei_mod:karras}
        \IfThenElse{$\ttl{range}_{left} = q$}
                   {$I_{q,left} \gets L_i$}
                   {$I_{q,left} \gets I_i$}             \label{l:apetrei_mod:left_child}\label{l:apetrei_mod:parent_update_begin}
        \If{$\ttl{range}_{right} = n-1$} \label{l:apetrei_mod:parent_rope_update_begin}
          \State $I_{q,skip} \gets \blacklozenge$
        \Else
          \State {$r \gets \ttl{range}_{right} + 1$}
          \IfThenElse{$\delta_{right} < \delta^*(r)$}
                     {$I_{q,skip} \gets L_r$}
                     {$I_{q,skip} \gets I_r$}             \label{l:apetrei_mod:internal_skip}
        \EndIf\label{l:apetrei_mod:parent_update_end}
        \State $i \gets q$                              \label{l:apetrei_mod:i_update}
      \Until{i = 0}
    \EndFor
  \end{algorithmic}
\end{algorithm*}